\documentstyle[prl,aps,multicol,epsfig]{revtex}
\begin{document} 
\draft 
\title{Energy of the nearest neighbor RVB state by systematic loop expansion}

\author{R. Eder}
\address{Institut f\"ur Theoretische Physik, Universit\"at W\"urzburg,
Am Hubland,  97074 W\"urzburg, Germany}
\date{\today}
\maketitle

\begin{abstract}
We present an approximation scheme for the calculation of the norm and 
energy of the nearest-neighbor-RVB state for the Heisenberg antiferromagnet 
on the 2D square lattice. The approximation leads to a systematic expansion 
of norm and energy, with the `expansion parameter' being the maximum length 
of loops taken into account in the calculation of energy and norm. The 
expansion converges well, the best estimate for the ground state energy/site 
is $-0.434473J$.
\end{abstract}
\begin{figure}
\epsfxsize=6.0cm
\vspace{-0.0cm}
\hspace{0.5cm}\epsfig{file=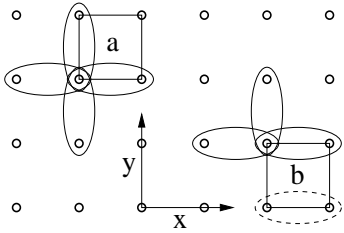,
height=3.0cm,width=4.5cm,angle=0.0}
\vspace{0.5cm}
\narrowtext
\caption[]{Interplay of loops and dimer coverings.}
\label{fig1} 
\end{figure}

 
\begin{multicols}{2}
The `RVB spin liquid' is a frequently occuring phrase in connection with
cuprate superconductors. It is supposed to describe a state
with strong short range singlet correlations, but no long range magnetic
order whatsoever. Despite its frequently being referred to in the literature,
up to now the `RVB spin liquid' is a rather elusive concept.
For example, there exists to date no simple and manageable
trial wavefunction for the $2D$ Heisenberg Antiferromagnet
which would describe such a state, nor is the precise
nature of its low lying elementary excitations known to any
degree of certainty
(as opposed e.g. to the antiferromagnetic state, where linear spin-wave
theory gives an excellent approximation to the ground state wave 
function and allows a quantitative discussion of its excitation spectrum).
Perhaps the best-defined `RVB spin liquid' is the
nearest neighbor RVB state\cite{Kivelson,Sutherland,Fradkin} 
- at least this wave function can be
written down in compact form.
We introduce the singlet generation operator on the bond  $i,j$ 
\[
s_{i,j}^\dagger = \frac{1}{\sqrt{2}}
(\;\hat{c}_{i,\uparrow}^\dagger
 \hat{c}_{j,\downarrow}^\dagger -
\hat{c}_{i,\downarrow}^\dagger
\hat{c}_{j,\uparrow}^\dagger\;),
\]
where $\hat{c}_{i,\sigma}^\dagger$$=$$c_{i,\sigma}^\dagger
c_{i,\bar{\sigma}}^{} c_{i,\bar{\sigma}}^\dagger$,
and the operator
\[
S = \sum_i ( s_{i,i+\hat{x}}^\dagger +  s_{i,i+\hat{y}}^\dagger),
\]
where $i+\hat{x}$ denotes the nearest neighbor of sites $i$
in $x$-direction.
The nearest neighbor RVB state on a 2D square lattice with $N$ sites
then is defined as
\begin{equation}
|\Psi\rangle = \frac{S^{N/2}}{(N/2)!} |vac\rangle
=\sum_{\phi} |\phi\rangle.
\label{RVB}
\end{equation}
In other words, the state consists of a superposition
of all possible distributions of $N/2$ singlets over the plane, $\phi$,
with equal phase. The total number of different singlet configurations
corresponds to the number of dimer coverings of the plane, $N_c$.
The calculation of $N_c$ is a well-known exactly solvable
problem from statistical mechanics\cite{Kasteleyn,Fisher} -
however, knowledge of this number is not really necessary to
estimate the ground state energy.
As will be seen below, the energy of the state (\ref{RVB}) is 
significantly higher than
that of the antiferromagnetically ordered ground state - 
the nearest-neighbor RVB state itself therefore is no candidate
for the ground state wave function. On the other hand, and this
is the main motivation for the present work, it has recently been
demonstrated for Heisenberg ladders\cite{Gopalan}
that by starting from
a `singlet vacuum' a theory for triplet-like fluctuations
`on top of' this vacuum can be constructed, which in fact compares very
well to numerical results. While for ladders
the topology of the system rather uniquely determines
one specific singlet covering,
the nearest-neighbor RVB state would be an appealing candidate
for a generalization of this approach to 
an isotropic and translationally invariant state of a planar system.
In trying to do so, however, one encounters an as yet prohibitive
technical obstacle: the fact that singlet configurations
corresponding to different dimer coverings are not 
orthogonal\cite{Sutherland}. 
More precisely, whenever we can draw a closed loop which
passes through both sites of each dimer it touches,
it is possible to shift the `train of dimers'
along the loop by one lattice spacing, so as to obtain a different
covering. Then, the `N\'eel components' in the
two dimer coverings along the loop are identical,
so that there is a nonvanishing
overlap. Each `N\'eel component' has the prefactor
$2^{-l/2}$,  where $l$ is the number of sites in the loop,
so that the overlap originating from the loop is
$2^{-(l/2-1)}$\cite{Sutherland}. Clearly the
most important contributions thus come from states which differ
only by short loops. First, the overlap due to short loops is
larger, and second, a long loop `blocks' more possibilities
for drawing other loops. In the following, we want to make these
considerations more quantitative.\\
To that end we write the norm as 
\begin{equation}
\langle \Psi | \Psi \rangle = 
N_c ( 1 + \overline{ \sum_{\phi \neq \psi}
\langle \phi | \psi \rangle } ),
\label{ndef}
\end{equation}
where $\bar{\;}$ denotes the statistical average over all configurations
$|\psi\rangle$.
As a first step we restrict the sum over $\phi$ to those
configurations which can be obtained from
$|\psi\rangle$ by `rotating' 
a certain number of $l$$=$$4$-loops. We denote the
number of these $4$-site loops by $\nu$, and
start with the case $\nu$$=$$1$. If there are $N$ sites in
the system, we can obviously draw $N$ different $4$-site loops
in the plane. However, in most cases the dimers of $|\psi\rangle$
will not `fit' into the drawn loop. Let us assume that we have
drawn a $4$-site loop somewhere in the plane, and
consider the lower left corner of the loop
(see the square labelled $a$ in Figure \ref{fig1}).
The probability that the dimer which covers this lattice 
\noindent 
site points into positive $x$ or positive $y$ direction, and thus `fits' 
into the loop, is $2/z$. Next, let us assume that the dimer on the lower 
left corner points in positive $x$ direction, and consider the upper
left corner (see the square labelled $b$ in Figure \ref{fig1}.
The dimer which covers this corner
cannot point into negative $y$ direction, because the
lower left corner is already occupied. It thus has
only $z-1$ possibilities for its direction.
If we now assume, that these $z-1$ orientations have
equal probability, the total probability that the
dimers in the state $|\psi\rangle$ fit into the prescribed
$4$-site loop is $\frac{2}{z(z-1)}$. 
Whereas for periodic boundary conditions
the first factor of $2/z$ follows rigorously from symmetry, the
second factor of $1/(z-1)$ is an approximation.
Numerical calculations on $4\times4$ and $6t\times 6$
lattices with periodic boundary conditions indicate
that the probability is higher, i.e. one should
replace $1/(z-1)$$=$$0.33333$$\rightarrow$$0.47$.
On the other hand the values of
$N_c$ obtained for these systems differ strongly from
the ones expected in the thermodynamical limit, so that the relevance
for the infinite system may be questionable. Since the difference is
not too large anyway, we stick to our simple estimate.
As will be seen below, a change of the probability
can be incorporated easily.\\
Next, if the dimers fit into the loop we can create a new state
$|\phi\rangle$ by rotating the dimers along the drawn loop,
and have $\langle \phi | \psi\rangle$$=$$1/2$.
The total contribution to the overlap from all states
which can be obtained by rotating one $4$-site loop in
$|\psi\rangle$ therefore is $N\lambda_1$ with
\[
\lambda_1 = \frac{1}{z(z-1)}= \frac{1}{12}.
\]
Next, let us assume that we have drawn
$\nu$ different $4$-site loops in the plane. The first question then
is: in how many different ways can we do that?
There are
$N$ ways to draw the first loop and since the second loop should not
intersect the first one, a number of positions for the
second loop are blocked. More precisely, the number of ways to
draw the second loop is only $N-b_{11}$, where
$b_{11}=9$. If the first two loops
are far from each other they will block $2b_{11}$ 
possible positions, so that the number of ways to draw the third loop is
$N-2b_{11}$. Continuing we estimate the
number $n_\nu$ of different ways to draw $\nu$ loops as
\begin{eqnarray}
&\;&
\frac{ N (N- b_{11})(N- 2 b_{11}) (N- 3 b_{11})\dots
(N- (\nu-1) b_{11})}{\nu !}
\nonumber \\
&=&
b_{11}^\nu
\left(
\begin{array}{c}
N/b_{11}\\
\nu
\end{array} \right).
\label{4loop}
\end{eqnarray}
The factor of $1/\nu!$ is due to the fact that by drawing
one loop after the other, each configuration of loops is generated
in $\nu!$ different ways. We have assumed that each loop
blocks precisely $b_{11}$ following loops, which is only
an approximation; deviations will occur if a loop is placed
`close' to a previously drawn loop. The probability for this to
happen is proportional to the `average density' of loops squared,
and we will have to check later on that this is small.\\
Assuming that (\ref{4loop}) is a reasonable approximation
for the number of different loop configurations,
we can now repeat the considerations for the single loop for each of the 
$\nu$ loops. In this way
we find that the total overlap of
$|\psi\rangle$ with all states which differ
by rotating $\nu$ $4$-site loops is simply
$n_\nu \lambda_1^\nu$, whence the contribution
to the overlap of $|\psi\rangle$ with {\em all} states which differ
by rotating an arbitrary number of $4$-site loops is
\begin{eqnarray}
N_4 &=& N_c \sum_{\nu>1}
\left(
\begin{array}{c}
N/b_{11}\\
\nu
\end{array} \right) (\lambda_1 b_{11})^\nu
\nonumber \\
&=& N_c\;[\;(1 + \lambda_1 b_{11} )^{N/b_{11}} -1\;].
\label{norm1}
\end{eqnarray}
The extra $1$ on the r.h.s. is negligible and will be dropped henceforth.
Note that we have extended the upper limit of the sum up to
$N/b_{11}$, where our above estimate certainly is invalid.
As mentioned above, our derivation will be valid only
if the `average density' of loops is small, and our next objective is
to check this assumption. To be more precise, we want to find out
which order $\nu$ in the expansion (\ref{norm1}) gives the
dominant contribution. Approximating the factorials
by Stirling's formula and nominally differentiating with respect
to the number of loops, $\nu$, we find that
the maximum contribution to the sum comes from terms with
\begin{equation}
\nu_0 = \frac{\lambda_1N}{1+ b_{11} \lambda_1} = \frac{N}{21}.
\label{plopp}
\end{equation}
Moreover, in the neighborhood
of $\nu_0$ we find $n_\nu  \lambda_1^{\nu}
\propto exp(-N (\nu-\nu_0)^2)$.
On the average, each drawn loop thus has $20$ `free squares'
around it, and terms which significantly different loop density give a 
negligible
contribution. This implies first that our approximation
of neglecting correlations between the loops is probably reasonably
justified for the terms which give the
dominant contribution to the norm, and second that extending the sum in
(\ref{norm1}) up to $N/b_{11}$ is justified as well.\\
Having obtained a first estimate for the norm of the
RVB state we proceed to compute
the expectation value of $H$ to the same
approximation. To that end we introduce the
projection operator ${\cal P}= |\psi\rangle \langle \psi|$
and ${\cal Q} = 1- {\cal P}$.
We find
\begin{eqnarray}
\langle \phi| H| \psi \rangle &=&
\langle \phi|({\cal P} + {\cal Q}) H| \psi \rangle
\nonumber \\
&=& \langle \phi|\psi \rangle E + \langle \phi|{\cal Q} H| \psi \rangle
\label{ham}
\end{eqnarray}
where $E$ denotes the expectation value of the
Hamiltonian in the state $|\psi\rangle$. We now write the expectation value
of the Hamiltonian as
\begin{equation}
\langle \Psi |H| \Psi \rangle = 
N_c ( -\frac{3NJ}{8} + \overline{ \sum_{\phi \neq \psi}
\langle \phi |H| \psi \rangle } ),
\label{hdef}
\end{equation}
and again restrict the summation to those states which can be
obtained from $|\psi\rangle$ by rotating an arbitrary number of $4$-site loops.
The first term on the r.h.s. in  (\ref{ham}) then simply gives the
energy of one dimer covering, $-3NJ/8$, multiplied by the approximate
norm. The second term is more involved and
represents the `true off-diagonal' contribution.
Let us assume that we act with the exchange term along a bond
$(i,j)$ connecting two sites covered by two different singlets in
$|\psi\rangle$. Then one has
\[
h_{ij}\; s_{i,i'}^\dagger s_{j,j'}^\dagger 
= \frac{J}{4} \sum_{\alpha=x,y,z} t_{i,i',\alpha}^\dagger 
t_{j,j',\alpha}^\dagger,
\]
where $t_{i,i',\alpha}^\dagger$ creates the $\alpha$-component
of the 3-vector of {\em triplets} on the bond $(i,i')$.
Next, a singlet and a triplet have opposite time reversal
parity, and a necessary condition for
a nonvanishing overlap along one loop is that
both states have equal time reversal parity
`along the loop'. We can conclude
that only those states $|\phi\rangle$ 
can have a nonvanishing matrix element with $|\psi\rangle$
which differ from $|\psi\rangle$ along a loop covering
{\em both} bonds, $(i,i')$ and $(j,j')$.
Let us assume that we have a dimer configuration 
\begin{figure}
\epsfxsize=6.0cm
\vspace{-0.0cm}
\hspace{2.0cm}\epsfig{file=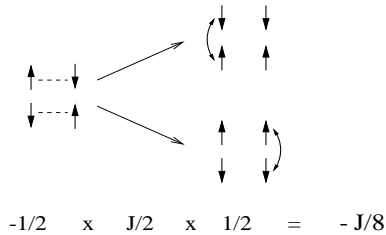,
height=3.0cm,width=5.0cm,angle=0.0}
\vspace{0.5cm}
\narrowtext
\caption[]{A contribution to
$h'$$=$$\langle \phi|{\cal Q} H| \psi \rangle$:
starting two singlets in $x$-direction (left state)
and acting with the transverse Heisenberg exchange 
along the indicated bonds one obtains two
states which are orthogonal to the initial state,
but correspond to singlets in $y$-direction.
Each of these channels gives a total
contribution of $-J/8$ to $h'$.}
\label{fig2} 
\end{figure}
\noindent 
which fits onto a $4$-site loop, and consider the matrix element
of $H$ between this configuration and the `rotated' dimer covering.
Looking at Figure \ref{fig2} one can read off
that the total contribution of this loop
to the matrix element is
\[
\langle \phi|{\cal Q} H| \psi \rangle = 
- 2\cdot 2 \cdot \frac{J}{8},
\]
where the additional factor of $2$ comes from
the `spin reversed' processes. Recalling that
the same configuration would have
contributed a factor of $1/2$ to the overlap of the
two states we find the total
off-diagonal contribution is given by
\begin{eqnarray*}
H_4 &=& - N_c J \sum_{\nu>1}
\left(
\begin{array}{c}
N/b_{11}\\
\nu
\end{array} \right) \; \nu \;(\lambda_1 b_{11})^\nu
\nonumber \\
&=& -\frac{NJ\lambda_1}{1+ \lambda_1b_{11}}
\end{eqnarray*}
This is just the expectation value
$\langle \phi|{\cal Q} H| \psi \rangle / \langle \phi|\psi \rangle$
for a single $4$-site loop
times the `most probable number' of loops, compare (\ref{plopp}).
Dividing by the norm we obtain our first estimate for the
expectation value of the energy:
\begin{equation}
E = - NJ\;[\; \frac{3}{8} + \frac{\lambda_1}{1+\lambda_1b_{11}}\;]
= -0.422619 NJ . 
\label{e1}
\end{equation}
Quite obviously, the correction due to off-diagonal
processes to the energy is rather small.\\
We proceed to the second step.
Whereas we restricted the sums in the norm and
expectation value of $H$ to configurations which differed only
by $4$-site loops from $|\psi\rangle$,
we now enlarge this
subset to comprise all states differing by $4$ {\em or} $6$-site loops.
Let us assume that we have drawn $\nu$ $4$-site loops in the plane,
and ask how many ways are there to draw $\nu_1$ additional $6$-site loops.
Putting the first $6$-site loop we have to make sure that it does not
intersect any of the $4$-site loops and assuming that each of the $4$-site loops
blocks a total of $b_{12}=15$ possible positions
of $6$-site loops, we find that the first $6$-site loop can be drawn in
$N_1$$=$$N - b_{12} \nu$ ways. Now, we proceed precisely as for the
$4$-site loops: every $6$-site loops that we draw blocks
$b_{22}=15.5$ other possible $6$-site loops (a $6$-site loop in $x$-direction
blocks $15$ $6$-site loops in $x$-direction and  $16$ $6$-site loops 
in $y$-direction), so that we estimate the number of ways
in which the $\nu_1$ $6$-site loops can be drawn as
\begin{eqnarray*}
&\;& \frac{N_1(N_1-b_{22})(N_1-2b_{22})\dots (N_1-(\nu_1-1)b_{22})}
{\nu_1!} 2^{\nu_1}
\nonumber \\
&=&
 (2b_{22})^{\nu_1}
\left(
\begin{array}{c}
(N- b_{12}\nu)/b_{11}\\
\nu_1
\end{array} \right).
\end{eqnarray*}
Thereby the extra factor of $2^{\nu_1}$ is due to the
fact that there are two different types of $6$-site loops
(in $x$-direction and $y$-direction).
By analogous considerations as for the $4$-site loop,
we now estimate that each drawn $6$-site loop gives a contribution of
\[
\lambda_2 = \frac{2}{4 z(z-1)^2} = \frac{1}{72}
\]
to the overlap. Proceeding as above we find that
the total contribution to the norm from all states
containing $\nu$ $4$-site loops and and arbitrary number of
$6$-site loops is
\begin{eqnarray*}
n_\nu \lambda_1^\nu&\;& (1 + 2\lambda_2 b_{22} )^{(N-b_{12}\nu)/b_{22}}
\nonumber \\
&=& n_\nu \lambda_1^\nu \;
(1 +2\lambda_2 b_{22})^{N/b_{22}} [\;(\frac{1}{1 +2\lambda_2 b_{22}})
^{b_{12}/b_{22}}\;]^{\nu}.
\end{eqnarray*}
Comparing this to the previous expression, $n_\nu \lambda_1^\nu$,
we note that the inclusion of the $6$-site loops has two effects:
a) an extra prefactor of $(1 +2\lambda_2 b_{22})^{N/b_{22}}$
which is independent
of $\nu$ and which will simply enter as an additional multiplicative
prefactor to the total norm, and b)
a renormalization of $\lambda_1$:
\[
\tilde{\lambda}_1= \lambda_1 (\frac{1}{1 +2\lambda_2 b_{22}})
^{b_{12}/b_{22}}.
\]
Neither modifications poses any problem within
our formalism and summing over $\nu$ we
obtain an improved estimate for the norm:
\[
N_6 = N_c\;
(1 +2\lambda_2 b_{22})^{N/b_{22}}\;
 (1 + \tilde{\lambda}_1 b_{11} )^{N/b_{11}}.
\]
To compute the energy to the same approximation, we note that
the contribution of a single $6$-site loop to the off-diagonal
matrix element $\langle \phi|{\cal Q} H| \psi \rangle$
is $-2J$\cite{Sutherland}. We thus find
the total off-diagonal energy of all states with
$\nu$ $4$-site loops and an arbitrary number of $6$-site loops is
\[
-[\;\nu J+\frac{4J(N-\nu b_{12})\lambda_2}
{1 + 2\lambda_2b_{22}}\;]
 (1 + 2\lambda_2b_{22})^{(N-b_{12}\nu)/b_{22}}\;]
n_\nu \lambda_1^\nu.
\]
Again, this replacement has two-fold effect:
there is a $\nu$-independent term of
$-(4NJ\lambda_2)/(1+2\lambda_2b_{22})$ times the total
overlap, which represents
the gain in energy due to inclusion of the $6$-site loops and enters
as an overall shift; second, there is a term
$\propto \nu$, which we can be reabsorbed into a renormalization of
$J$ in the off-diagonal energy due to $4$-site loops:
\[
\tilde{J} = J\;(\;1 -\frac{4b_{12}\lambda_2}{1 + 2\lambda_2b_{22} }\;).
\]
This takes into account that drawing
a $4$-site loop the number of possible $6$-site loops is reduced
resulting in a reduction of the energy gain
from a single $4$-site loop. 
Summing again over $\nu$ and dividing by the
norm we obtain an improved second estimate for the energy:
\begin{eqnarray}
E &=& - N\;[\; \frac{3J}{8} + \frac{\tilde{\lambda}_1 \tilde{J}}
{1+\tilde{\lambda}_1b_{11}} +\frac{4J\lambda_2}
{1 + 2\lambda_2 b_{22}}\;]
\nonumber \\
&=& -0.435338 NJ. 
\label{e2}
\end{eqnarray}
The `most probable number' of $4$-site loops drops
by $\approx 15$\%
from $N/21$$=$$0.0476N$ to $0.0403N$, whereas the most probable
number of $6$ loops is $0.0194N$.
Enlarging the subset of states used for the computation of
energy and norm does have only a moderate influence, i.e. 
our procedure is reasonably convergent.\\
In addition the procedure has an induction-like
character: enlarging the sums in (\ref{ndef}) and (\ref{hdef})
the subset of states
leads to additional prefactors in the norm,
a renormalization of the preceding $\lambda$'s, additional
terms in the expectation values of $H$
and a renormalization of the exchange integrals in the
preceding energy contributions. Repeating everything
including also $8$-site loops gives an estimate for the
energy of $E = -0.434473 NJ$,
the number of $4$-loops drops by only $8$\%.
The convergence thus is quite fast, with the
difference between subsequent estimates dropping by an order of magnitude in
each step (actually the dependence is not monotonous in that the
estimate including $8$-loops is higher than the
the one from $6$-site loops). We can conclude that taking into account
only $4$-site loops may be sufficient
to get reasonably accurate results with little effort - this may be an 
important simplification in further applications of the
state (\ref{RVB}).\\
Comparing with results of numerical calculations\cite{Liang,Havilio}
which give an energy of $\approx -0.6J$/site the agreement is obviously
not too good. This is probably due to thefact that our simpel estimates
for the probability to find a given loop in an arbitrary dimer 
configuration are not too accurate.
 
\end{multicols}

\begin{references}
\bibitem{Kasteleyn}
P.W. Kasteleyn, Physica {\bf 27} (1961) 1209.
\bibitem{Fisher}
M.E. Fisher, Phys. Rev. {\bf 124} (1961).
\bibitem{Kivelson}
 S. Kivelson, D. Rokhsar, and J. P. Sethna,
 Phys. Rev. B {\bf 35}, 865 (1987).
\bibitem{Sutherland} 
 B. Sutherland, Phys. Rev. B {\bf 37}, 3786 (1988).
\bibitem{Fradkin}
 E. Fradkin, {\em Field Theories of Condensed Matter Systems},
Addison-Wesley Publishing Company, 1991.
\bibitem{Gopalan}
S. Gopalan, T. M. Rice, and M. Sigrist,
Phys.\ Rev.\ B {\bf 49}, 8901 (1994);
see also R. Eder, cond-mat/9712003.
\bibitem{Liang}
 S. Liang, B. Doucot, and P. W. Anderson
 Phys. Rev. Lett. {\bf 61}, 365 (1988).
\bibitem{Havilio}
 M. Havilio, Phys.\ Rev.\ B {\bf 54}, 11929 (1996).
\end{references}
\end{document}